# Introducing a response-based duration metric and its correlation with structural damages

Mohammadreza Mashayekhi[1]*, Mojtaba Harati[2]*, Morteza Ashoori Barmchi[1] and Homayoon E. Estekanchi[1]

[1]Department of Civil Engineering, Sharif University of Technology, Tehran, Iran
[2]Department of Civil Engineering, University of Science and Culture, Rasht, Iran

**Abstract**

This study proposes a response-based parameter for strong motion duration which is computed for structures and is the total time they are nonlinear during an earthquake. Correlation between structural response and duration for structures, subjected to a set of spectrum matched ground motions, is employed to examine the efficiency of the proposed method. The spectral matching procedure ensures that the influence of amplitude and frequency content of motions on structural response variability is significantly removed. Four concrete building type systems are studied and correlation coefficients of structural response with the proposed duration definition are examined. Comparison of the proposed method with other existing definitions—the record-based and response-based metrics—shows about 15-20% improvement in the correlation values.

## 1 Introduction

Duration along with amplitude and frequency content are three main characteristics of ground motions that must be identified to characterize an earthquake record. Frequency content and amplitude are reflected in the acceleration spectrum and are presently considered in the record selection procedure of current building codes (e.g. ASCE/SEI 41-17 (2017) and FEMA-356 (2000)). There is no doubt in regard to the influence of amplitude and frequency content in structural responses while the impact of motion duration on structural responses is still a matter of debate (Hancock and Bommer 2006). Earthquake duration length can play an important role in the structural responses. Several investigators have addressed the influence of motion duration on the seismic response of different structures. These studies revealed that seismic responses of the structures under earthquake loadings with deteriorative behaviors, including RC frames (Belejo et al. 2017; Hancock and Bommer 2007), concrete dams (Wang et al. 2015; Bin Xua et al. 2018) and masonry buildings (Bommer et al. 2004), are directly influenced by duration of ground motions. It implies that structures with deteriorating behaviors are much more susceptible to motion duration (Mashayekhi and Estekanchi (2012, 2013)). In this case, accumulated damage indices which are partially or completely based on the hysteretic cyclic energy of the earthquakes, such as Pak-Ang damage index (Park and Ang 1985), are shown to have higher positive correlations with the motion durations. However, the extreme damage indices such as peak floor drifts or peak plastic rotations of the elements are demonstrated to be almost uncorrelated to the motion duration (Hancock and Bommer 2007; Mashayekhi et al. 2019). It is of the essence to mention that the same results likewise apply for the steel (Bravo-Haro and Elghazouli 2018) and wood frame (Pan et al. 2018) structures. Guo et al. (2018) also demonstrated that duration of near-fault pulse-like ground motions has also a significant positive correlation with the earthquake-induced structural demands of such motions.

*These authors have contributed equally to the work
Corresponding: Mohammadreza Mashayekhi, Research Associate, Sharif University of Technology, Tehran, Iran. Email: mmashayekhi67@gmail.com

Given the fact that motion duration can have a substantial influence on the structural responses, definitions of earthquake duration obviously need to be well quantified to show the possible existing relationship between motion duration and the potential destructiveness power of the earthquakes. Hence, in order to characterize this intensity measure of the earthquake, researchers are working on the topic of motion duration definition since 1962 through a pioneer study by Rosenblueth and Bustamante (1962). All the present definitions of strong ground motion duration can be divided into two distinct groups, including the record-based and response-based definitions of motion duration (Bommer and Martinez-Periera 1999)

The record-based definitions are principally based on characteristics of ground motion records. There are numerous definitions for motion duration in the literature, but some of them are more commonly accepted and used by the earthquake engineering community. These definitions are of the bracketed-, uniform- and significant-type metrics for motion duration. The bracketed duration of motion delivers the total time left between the first and last acceleration excursions which are greater than a specific predefined threshold. This threshold can be of an absolute (0.05g or 0.1g) or a relative kind and is selected in a way that it is believed it can cause damages to the structure of interest. The definition pertinent to the uniform duration is all related to the sum of the elapsed time intervals considering the same aforementioned threshold level set on the acceleration time series of motions. But the definition related to significant duration is somehow different from the bracketed and uniform duration, which makes use of a well-known integration-based accumulative intensity measure, the so-called Arias Intensity (AI). Significant duration is denoted by $D_{x-y}$ hereafter, which is defined as the time interval during which the normalized AI moves from a minimum (x%) to a maximum (y%) threshold. And so, the $D_{5-95}$ means the time interval as buildup accumulation energy of the earthquake goes up from 5 to 95 percent (Trifunac and Brady 1975). Moreover, new duration definitions are also proposed in order to marginally improve the correlation of motion duration with structural damage, which can be developed from the existing record-based metrics that are combined (Taflampas et al. 2009) or modified (Bommer and Martinez-Periera 1999; Rupakhety and Sigbjörnsson 2014) to make new definitions for motion duration.

On the other hand, and contrary to the record-based definitions, the response-based metrics for motion duration are expected to be more pertinent to the seismic responses of the structures. These duration-related intensity measures are normally based on the definitions whose parameters are extracted from seismic characteristics or SDOF models that are reasonably representative of the structures being evaluated. Rosenblueth and Bustamante (1962) were the first researchers who proposed a rather complicated—and not easily usable—response-based definition. It was the duration of an equivalent motion with uniform intensity per time, which was primarily defined to study the influence of structural damping on the spectral ordinates of ground motions. This uniform motion is required to generate a given ratio between the maximum spectral displacements of two linear SDOF systems with the same predefined period of vibration, one SDOF without damping and the other one with a specific damping ratio.

Response-based definitions for motion duration are occasionally in line with those general concepts described for the record-based definitions—the bracketed, uniform and significant duration concept. For instance, Perez (1980) proposes a structural response definition which is somehow of the uniform-type concept. This motion duration is the whole time during which the velocity response in the time history of an elastic SDOF is above a specified threshold. However, it is worthwhile to mention that the duration definitions offered by Perez (1980) as well as Rosenblueth and Bustamante (1962) are totally defined on an elastic or linear SDOF system and are not directly related to ground motion acceleration which is shown to have good correlations with earthquake damages.

The concept of uniform-type motion duration also inspired Xie and Zhange (1988) to give a new response-based definition, but they put forward a quantifiable threshold for this definition of motion duration—the yield acceleration. Contrary to the constant values of threshold used in the uniform duration, e.g. 0.05g or 0.1g, the yield acceleration is however a function of seismic characteristics of the structure, including the yield strength level in a bilinear system, total mass as well as the ordinate of the response spectrum of the acceleration with damping ratio and natural period of the structures being investigated. It is worth adding that he applied this definition of motion duration on a number of nonlinear SDOF systems and found that there is a high positive correlation between this response-based motion duration and structural collapse of the selected models. Zahrah and Hall (1984) reached a response-based duration definition through nonlinear time history analysis performed on several bilinear SDOF systems (or SDOFs with zero post-yield stiffness) with different natural periods. Their definition is quite similar to the significant-type duration, being the length of time during which 5 to 75 percent of the earthquake energy in a structure is dissipated inelastically.

This study proposes a new parameter for the duration of ground motions, which is based on the nonlinear response of structures. In order to evaluate the efficiency of the proposed definition, the correlation of the induced damages to the selected structures, which are exposed to spectrally matched ground motions, with the duration of earthquakes is investigated. It is assumed that a spectral matching procedure removes the variability associated with amplitude and frequency content of ground motions. Four concrete structures are considered in this study. Correlation of structural damages with the proposed duration definition is computed and compared with the existing duration definitions—the bracketed, uniform, significant and two other relevant response-based duration definitions.

## 2 Proposed definition

The definition presented here is based on the nonlinear response of an equivalent single degree of freedom structure which is subjected to a ground motion of interest. In fact, the proposed duration parameter is calculated for a particular structure. This metric which is named Inelastic Duration (IDU) can measure the total time that the equivalent SDOF structure, as shown in Figure 1, is nonlinear during the excitation. In this figure, $K_e$ is the equivalent linear stiffness of the structure which is calculated based on the natural period of the structure being investigated, the total mass ($m$) of the structural system and $\zeta$ which is the selected damping ratio. The equivalent strength level of the SDOF system ($Fy_{eqv}$) is 60 percent of the strength level computed for the whole structural system, also known as $Fy$ in the literature. The $Fy$ is and can be routinely derived by a pushover analysis. The equivalent strength level corresponds with the formation of the first plastic hinge of the structure (ASCE/SEI 41-17 (2017); FEMA-356 (2000)), but the strength level obtained through a pushover procedure is related to the yielding of the whole structure. By considering $Fy_{eqv}$ instead of $Fy$, all time instants of the motion—which have the potential to cause at least one plastic hinge—are counted for the proposed strong motion duration. The proposed duration is defined by Equation (1):

$$t_{eqv} = \int_0^{t_{max}} \left(\frac{K_e - K_T(t)}{K_e}\right) dt \quad (1)$$

where $K_e = \dfrac{4\pi^2 m}{T^2}$

where $K_T$ is the tangential stiffness of the equivalent SDOF system and $t_{max}$ is the total duration of the ground motion. The Newmark-beta method (Newmark 1959) is employed for the dynamic analysis.

In the view of Equation (1), $t_{eqv}$ depends on characteristics of the equivalent SDOF system; SDOF systems possess two main attributes: natural period and strength level, which are denoted by T and $Fy_{eqv}$ ($= 0.6Fy$), respectively. To model equivalent SDOFs, different $0.6F_y/W$ parameters and natural periods of the candidate MDOF systems are taken to be applied as indicated in Table 2 (section 4.1), where $W$ is the total seismic weight of the MDOF structural models and $F_y$ stands for their strength levels. The schematic illustration of the equivalent SDOF system is shown in Figure 1.

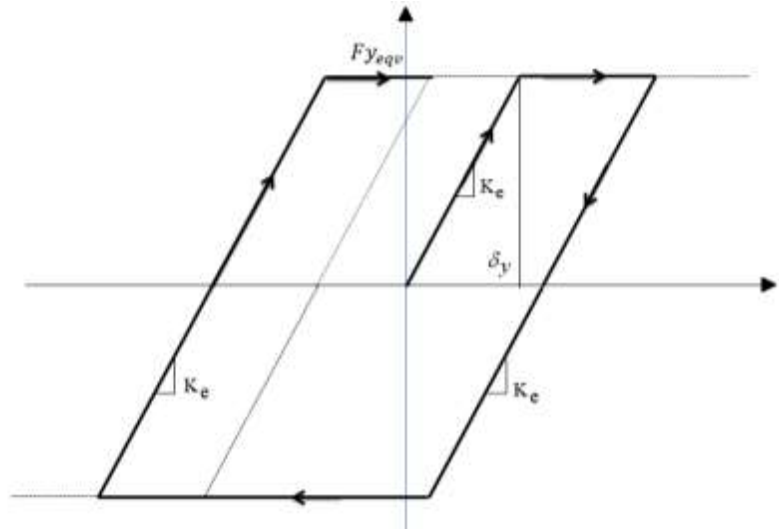

Figure 1. Schematic illustration of the equivalent SDOF system

As mentioned before, a static pushover procedure can be used to obtain $F_y$ for each structural system of interest, including steel, concrete and wood building type systems. Also, based on the expected displacement ductilities, the strength level of structures can be approximately found using R-μ-T functions (e.g. Miranda and Bertero (1994); Nassar and Krawinkler (1991)) as suggested by Aschheim and Black (2000). Since strength levels of the structures can be appropriately estimated at different structural periods, a rather new concept named duration spectrum can be defined for motion duration. Duration spectrum is a plot of motion duration against the natural period of the structural system being considered. The duration spectra are plotted for structures with different periods of vibration but for the same ductility ratio. Using an R-μ-T relationship offered by Miranda and Bertero (1994), a duration spectrum sample for a ground motion related to the Loma-Prieta earthquake of 1989 is computed and presented in Figure 2. As can be seen from this figure, strong motion duration or spectral duration of a specific earthquake gets increased in general when built infrastructures are expected to experience more nonlinearity, i.e., structural systems with ductility ratio equal to 4 or even more.

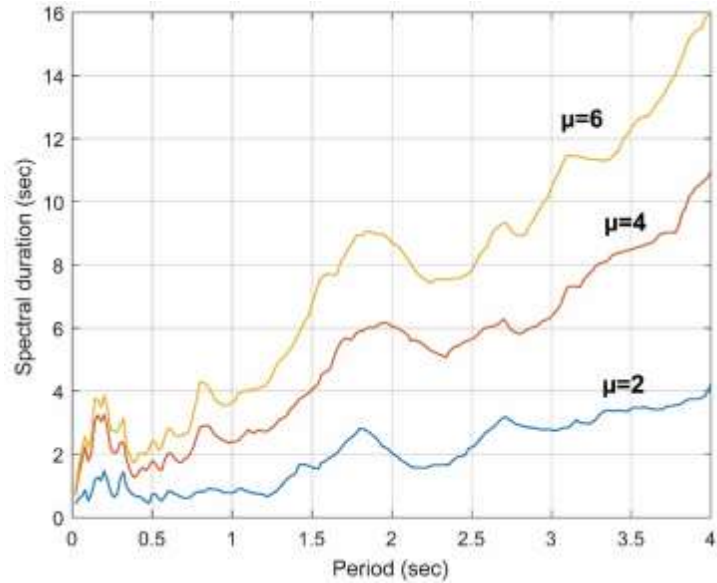

Figure 2. Duration spectra computed for the Loma-Prieta earthquake of 1989 at different expected displacement ductility ratios

## 3 Adopted ground motions and spectral matching procedure

### 3.1 Characteristics of the selected motions

In this study, 200 ground motions selected by Heo et al. (2011) are taken as a source for the record selection procedure. Heo et al. (2011) used these motions in the dynamic analysis which aimed to compare amplitude scaled and spectrum-matched ground motions for seismic performance assessment. In their study, unscaled ground motions that drive the structure into the nonlinear range was of interest, therefore, a subset of 200 ground motions from Pacific Earthquake Engineering Research (PEER) Next Generation Attenuations (NGA), whose PGA exceeded 0.2g was used in their computer simulation. The magnitude-distance distribution of the selected earthquake records is extracted from Heo (2009) and plotted in Figure 3. It contains information related to the two components of 100 earthquakes. Soil type C and D denotes to the soft and deep stiff soils, respectively.

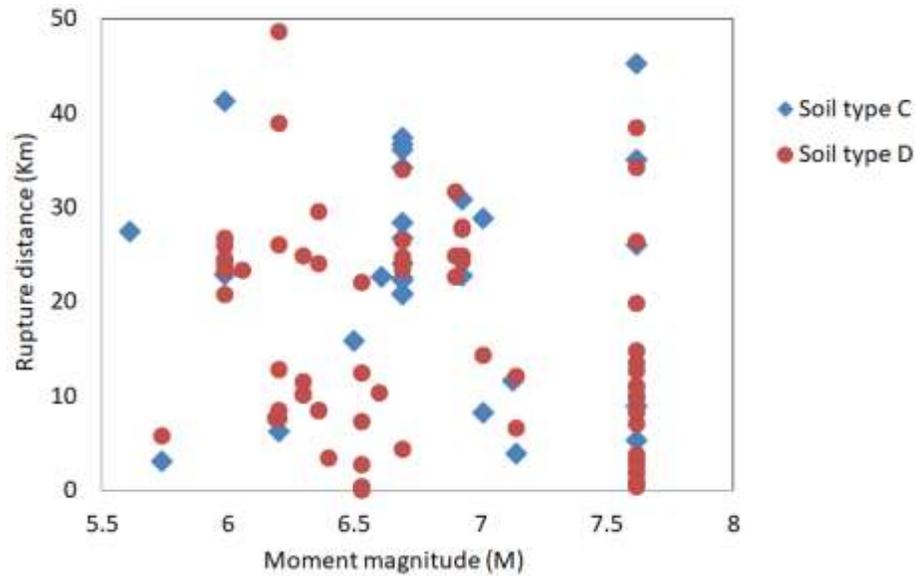

Figure 3. Distribution of magnitude-distance of the dataset

Acceleration, as well as the duration spectra of these ground motions, are shown in Figure 4 (a) and (b). Duration spectrum of these ground motions is computed for a ductility ratio equal to 4. This ductility level can be expected to occur in structures prepared for a design basis earthquake (DBE). As can be seen in Figure 4, dispersion of acceleration spectra of these ground motions is very noticeable which can also cause a considerable dispersion in the structural responses. This variability can be attributed to the amplitude and frequency content of the ground motions. As far as the effect of motion duration is concerned, this source of variability should be minimized. In order to reduce this type of variability, acceleration spectra of the ground motions are matched with a target spectrum.

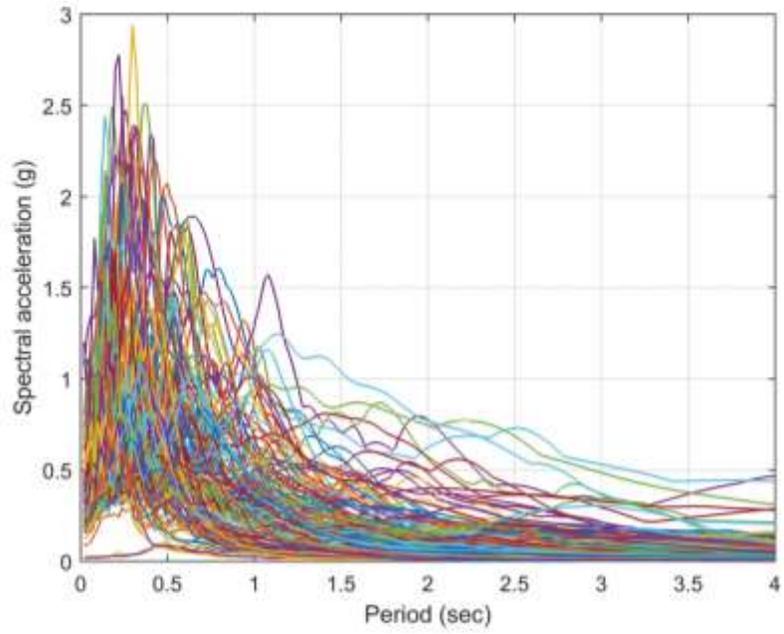

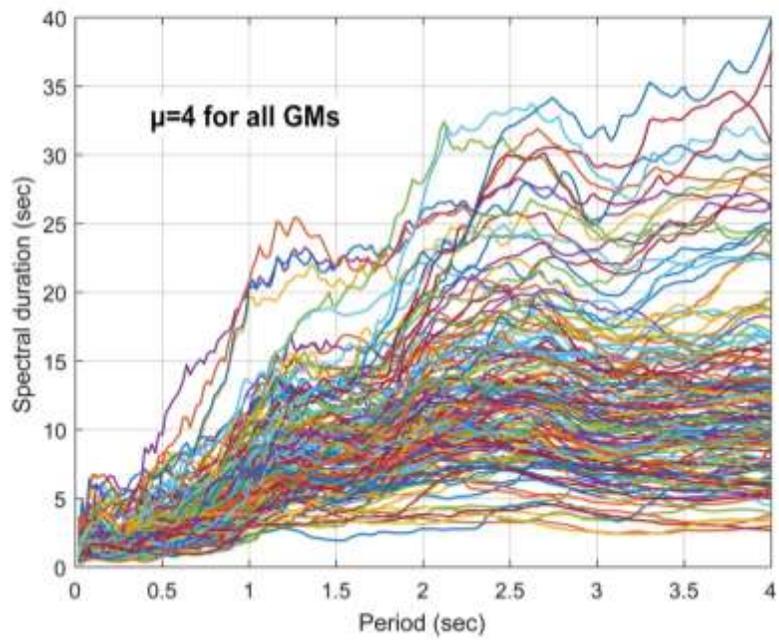

Figure 4. Acceleration and duration spectra of selected ground motions

### 3.2 Spectral matching procedure

To remove and diminish the influence of spectral amplitudes of ground motions from the characteristics of the selected motions, earthquakes are matched to a target response spectrum using the spectral matching procedure developed by Hancock et al. (2006). Hence, all of these matched motions only differ in terms of motion duration and non-stationary characteristics. It is important to note that the matched records inherit these two aforementioned characteristics from the original earthquake ground motions.

In this study, the target spectrum is the median of the spectra of the selected ground motions. The comparison of acceleration spectra of the original and matched time histories together with the associated target spectra, acceleration and displacement target spectra extracted from the original time series, are also presented in Figure 5 and Figure 6, respectively. Figure 6 indicates that the ground motions are well matched to the target spectra with a minimal change seen in the initial time histories of the ground motions. Acceleration spectra of matched ground motions versus the respective target spectrum, as shown in Figure 6 (a), demonstrate an acceptable match.

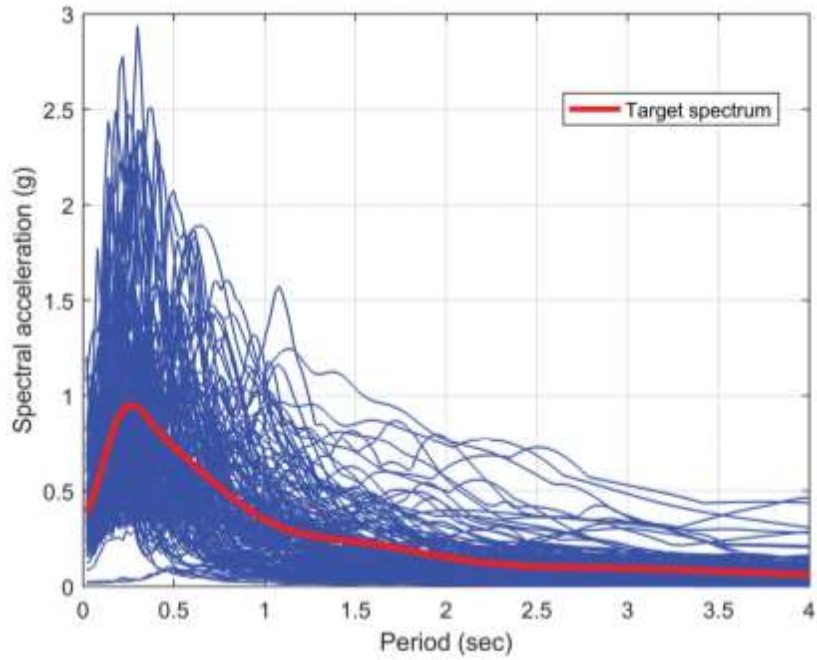

(a)

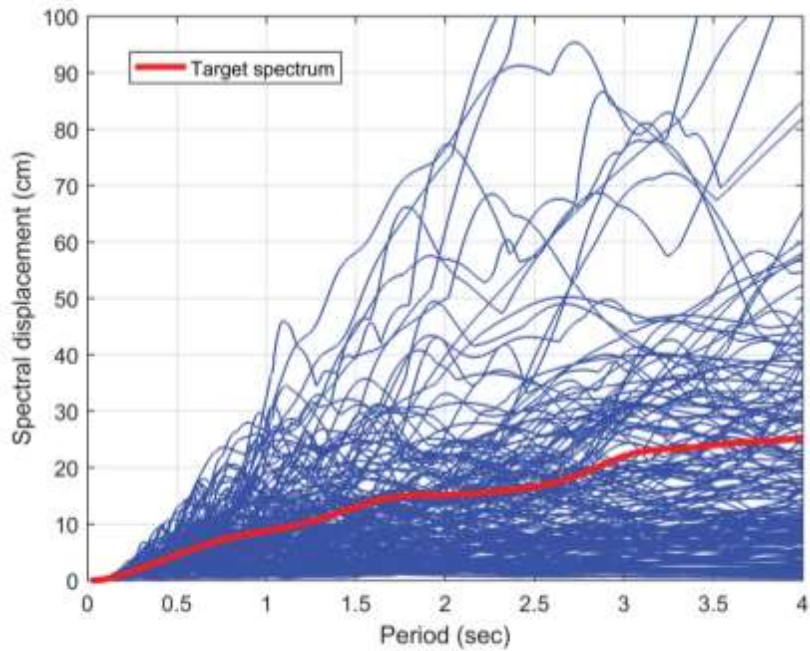

(b)

Figure 5. Spectral-related characteristics of the selected ground motions: a) acceleration spectra versus the target spectrum; b) displacement spectra and displacement target spectrum

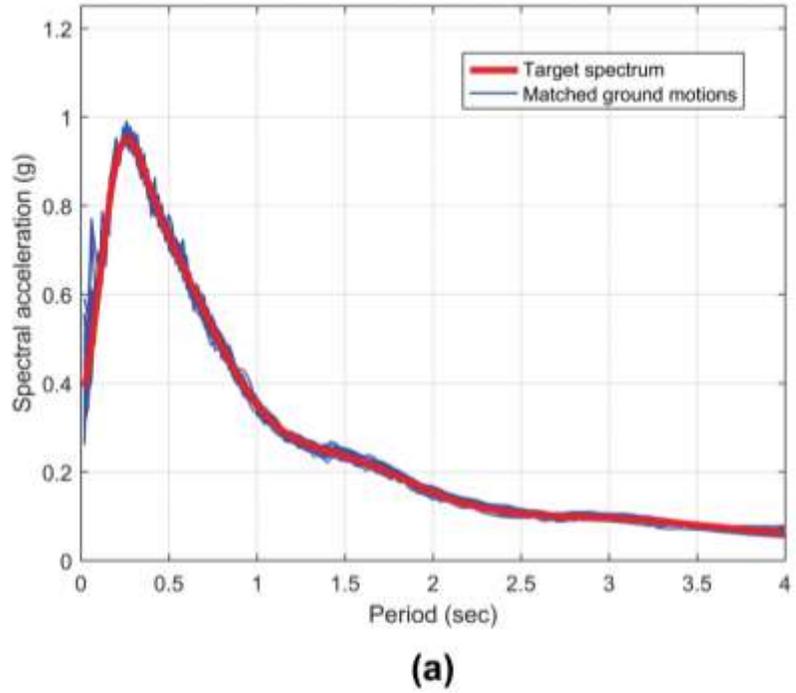

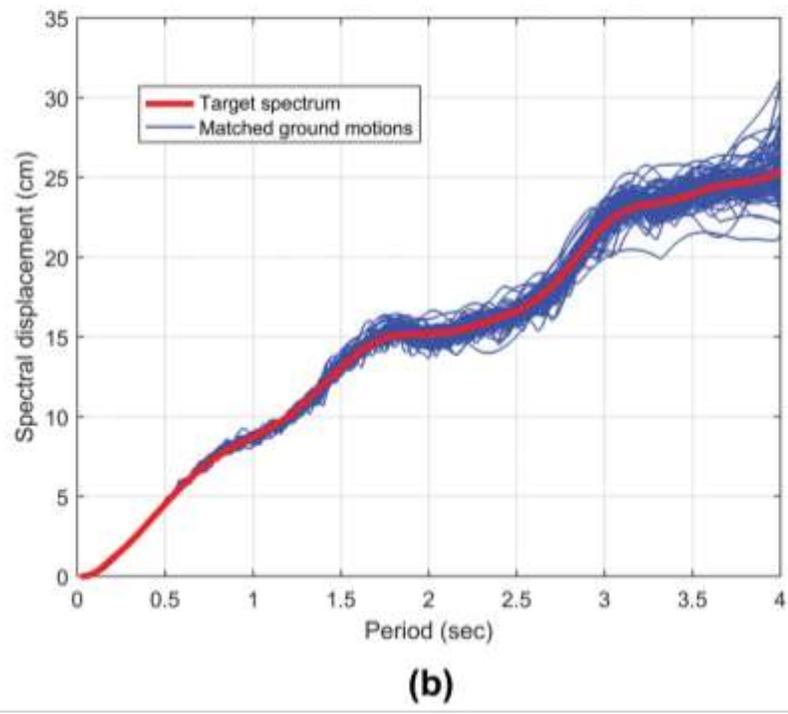

Figure 6. Spectral-related characteristics of adjusted ground motions: a) matched acceleration spectra versus the target spectrum; b) matched displacement spectra and displacement target spectrum

# 4 Structural modeling and considered damage indices

## 4.1 Structural models

The equivalent SDOF systems used for the proposed duration definition are created and modeled based on the bilinear pushover curves of several 2-D building type frames. These RC structures are adopted from Korkmaz and Aktaş (2006) and their general characteristics including stories numbers, bay width, height length, and the total height of the structures are provided in Table 1. Figure 7 represents the general configuration of these selected structures.

Table 1 Essential specifications of the selected RC structures

| Model ID | Number of Stories | Typical Bay Lengths (m) | Typical Story Height (m) | Total Structural Height (m) |
|---|---|---|---|---|
| 1003 | 3 | 6,4,6 | 3.0 | 9.6 |
| 1005 | 5 | 6,4,6 | 3.0 | 15.6 |
| 1008 | 8 | 6,4,6 | 3.0 | 24.6 |
| 1012 | 12 | 6,4,6 | 3.0 | 36.6 |

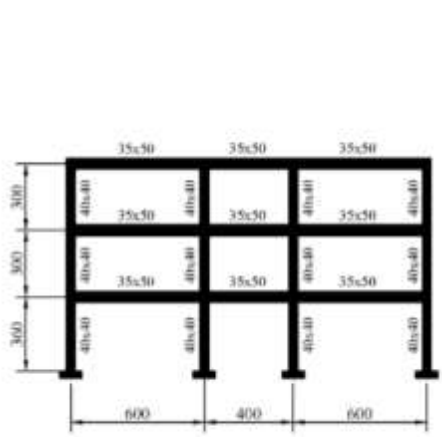
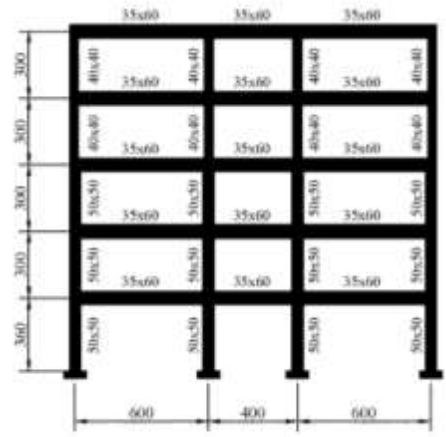
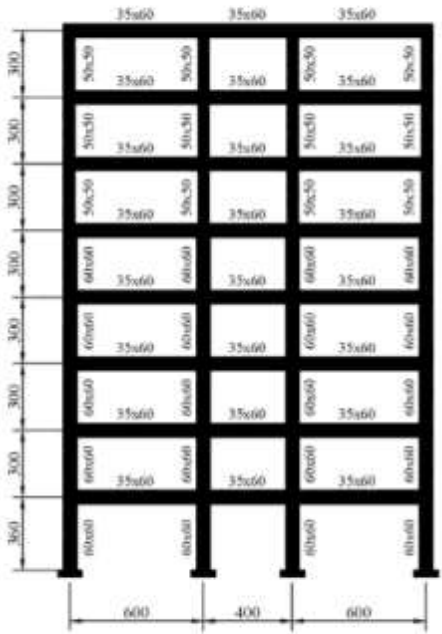
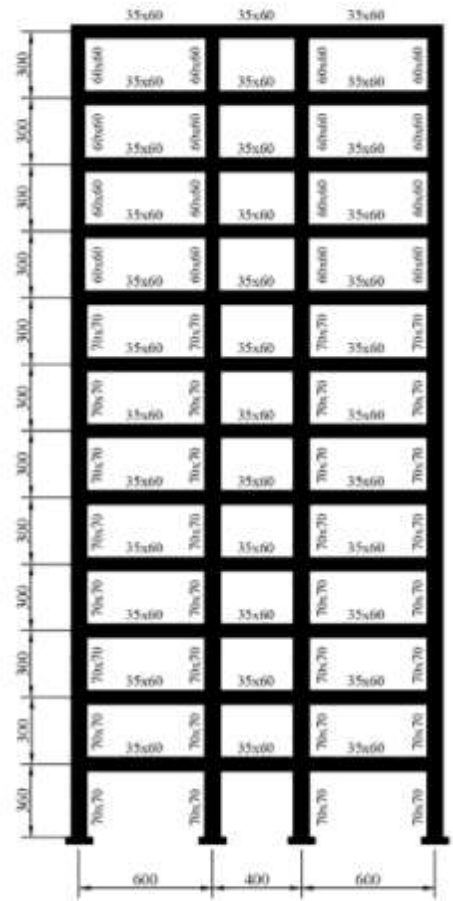

Figure 7. Illustration of the selected structures (Karimzada 2015): a) a 3-story structure; b) a 5-story structure; c) an 8-story structure; d) a 12-story structure

These structures are numerically modeled by fiber section method in Opensees software (McKenna 2014). This software is an open-source platform which has been developed in UC Berkeley for any kinds of dynamic analysis. Inelastic frame element with concentrated plasticity is employed to model beams and columns. Also, nonlinear input characteristics—known as Concrete01 and Steel02—are used for material modeling, which are correspondingly selected for concrete and steel fibers, respectively. Note that in this nonlinear modeling procedure floors are assumed to behave as rigid diaphragms, so the rigid constraints are activated using the equalDOF command of the Opensees. Expectedly, to avoid flaws in the process of simulation and to guarantee realistic modeling, a 5% tangent stiffness proportional damping (Priestley and Grant 2005; Zareian and Medina 2010) is also employed throughout this study.

Equivalent SDOFs of these structures are modeled by converting their pushover curves to bilinear curves. Figure 8 depicts this procedure for model 1008. Table 2 lists parameters associated with the equivalent SDOF of the considered structures.

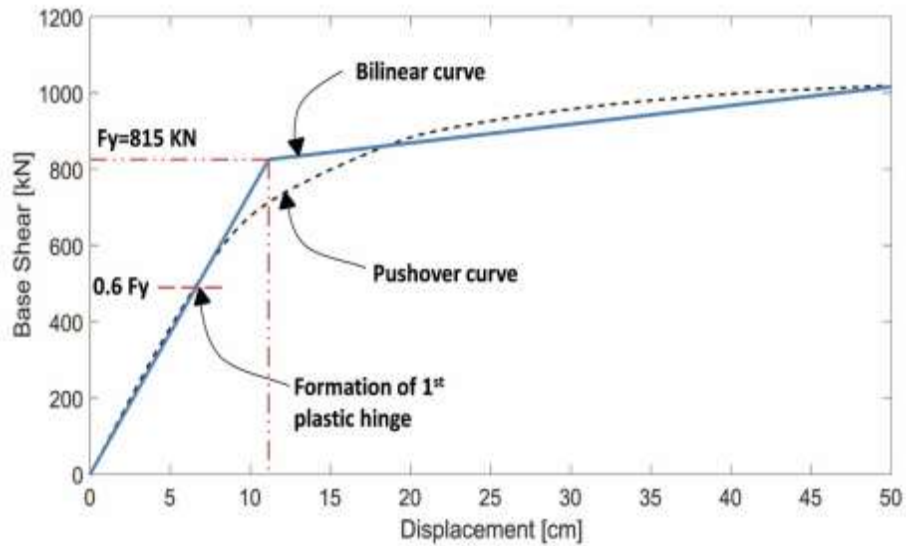

Figure 8. A pushover curve for model 1008 and its bilinear equivalent to model the relevant equivalent SDOF

Table 2 Characteristics of bilinear pushover curves required to create the equivalent SDOF models

| Model ID | Fy (KN) | Total weight of structure, W (MN) | 0.6×(Fy/W) | Fundamental Period, T1 (sec) |
|---|---|---|---|---|
| 1003 | 500 | 2.04 | 0.144 | 0.61 |
| 1005 | 800 | 3.4 | 0.138 | 0.69 |
| 1008 | 815 | 5.24 | 0.096 | 1.22 |
| 1012 | 1100 | 8.16 | 0.082 | 1.4 |

.

### 4.2 Damage indices

Damage induced in a structure, in general, is represented as a combination of its maximum response and its absorbed hysteretic energy. Based on different combinations of maximum response and hysteretic energy, several damage index definitions are available in the literature (Kappos 2005). As one of the most well-known definition, the Park-Ang damage index (Park and Ang 1985) brought in Equation (5) is employed in this study.

$$DI_m = \frac{\theta_m - \theta_r}{\theta_u - \theta_r} + \frac{\beta}{M_y \theta_u} \int M \, d\theta \tag{5}$$

where $\theta_m$ is the maximum rotation recorded during the time history analysis; $\theta_u$ is the ultimate rotation capacity of the section, which is computed based on the recommendation offered by FEMA P695 (2009); $\theta_r$ is the recoverable rotation of the section when unloading occurs, and it is calculated based on an equation derived by Fardis and Panagiotakos (2002); $M_y$ is the yield moment that is according to the equations derived using section analysis as described by Biskinis and Fardis (2009); $\int M \, d\theta$ is shown by term EHm in this paper, which is the cumulative energy absorbed within the member section; and the β factor—which is set to 0.05 as recommended by Park et al. (1987) for the members of an RC frame—is an empirical parameter altering the balance between the extreme displacement response and the energy term of the section, $\int M \, d\theta$.

The Park-Ang damage index may be expressed locally, for an individual element, or globally, either for a single story or for the whole structure. In the former one, the maximum value of $DI_{ms}$ calculated for two end sections of the member are considered as the member damage index value. The Park-Ang damage index for a particular story, denoted by $DI_s$-I, and the one for the whole structural level ($DI_{total}$) are computed based on the relationships suggested by Reinhorn et al. (2009), which use weighting factors depending on the hysteretic energy dissipated at the component and story levels, respectively.

As mentioned before, several engineering demand parameters—or earthquake damage measures—are considered for the correlation computations. These measures include the total Park-Ang damage index, the member and total hysteretic energy of the structure, Park-Ang damage index of beams and columns located at the first story level (or ground floor). Also, the Park-Ang damage measure for the first story level of considered structural models is incorporated as a demand measure in this study.

## 5    Numerical Results

In this section, numerical results at design-based hazard level (or at DBE level) are presented. For each structure, the spectrum-matched ground motions are linearly scaled so that their acceleration spectra at the structure's first mode period become equal to the DBE acceleration spectrum. A 2/3 of the acceleration spectrum proposed by ASCE07 (2010) for Los Angeles is taken as DBE acceleration spectrum. The DBE hazard level has the exceedance probability of 10% in 50 years.

As illustrated in Figure 9, the proposed duration measure against the total damage index and total hysteretic energy are considered here for the analysis at the DBE level in the model 1012. It can be seen that the total damage measures, based on the Park-Ang index and energy hysteresis of the whole system, have a high correlation with the duration definition proposed in this study. As employed by other researchers (Guo et al. 2018; Han et al. 2017; Hancock and Bommer 2007), the Pearson linear correlation coefficient is utilized hereafter as an indicator to check the efficiency of the proposed duration measure. Since the influence of amplitude-based characteristics of the original ground motions is adequately excluded in the adjusted seismic inputs, the value of linear correlation coefficient found between earthquake duration and the applied damage measures may not be far from the exact value supposed to exist for such quantity.

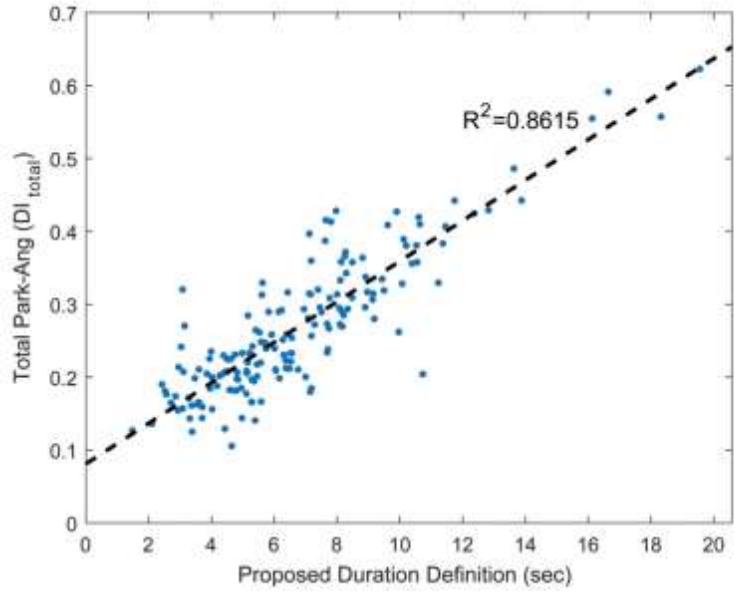

(a)

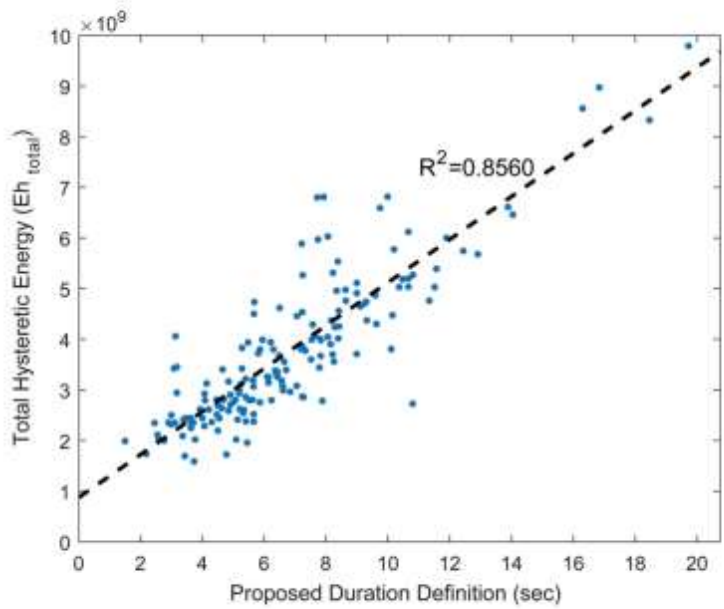

(b)

Figure 9. Correlation of motion duration and damage indices for model 1012: a) proposed duration definition versus the total damage of the system; b) proposed duration definition versus the total hysteretic energy of the structure

## 6  Comparison with existing duration metrics

Table 3 provides correlation coefficient values of duration and response parameter for model 1012 by using the proposed metric (or inelastic duration) and three existing record-based definitions, the bracketed (Br), uniform (U) and significant duration (Dx-y) metrics. Both relative (5–95% and 5-75% significant duration) and absolute (0.05g or 0.1g thresholds for both uniform and bracketed types) definitions are used for the comparison purposes in this regard. It is of the essence to note that the aforementioned record-based definitions are commonly accepted within the earthquake engineering community and are frequently used as metrics to measure this characteristic of the ground motions. Results show that the inelastic duration (IDU) may have up to 10% more correlation than the existing records-based definitions for a 12-story RC frame.

Table 3 Correlation coefficients of response parameters with the record-based duration metrics for a 12-story structure.

| Response parameter | Correlation Coefficients | | | | | | |
|---|---|---|---|---|---|---|---|
| | Proposed definition | Bracketed 0.05g | Bracketed 0.1g | Uniform 0.05g | Uniform 0.1g | $D_{5-75}$ | $D_{5-95}$ |
| $DI_{total}$ | 0.862 | 0.791 | 0.774 | 0.825 | 0.758 | 0.762 | 0.820 |
| $DI_{s-1}$ (1$^{st}$ story) | 0.809 | 0.658 | 0.662 | 0.709 | 0.666 | 0.662 | 0.695 |
| $DI_m$ (beam) | 0.758 | 0.635 | 0.641 | 0.679 | 0.631 | 0.629 | 0.678 |
| $DI_m$ (column) | 0.819 | 0.675 | 0.677 | 0.726 | 0.684 | 0.679 | 0.707 |
| $EH_{total}$ | 0.856 | 0.817 | 0.800 | 0.868 | 0.818 | 0.807 | 0.831 |
| $EH_m$ (beam) | 0.867 | 0.779 | 0.755 | 0.805 | 0.740 | 0.745 | 0.804 |
| $EH_m$ (column) | 0.856 | 0.730 | 0.714 | 0.775 | 0.731 | 0.724 | 0.756 |

The same procedure is performed for the other three structures. Results obtained from these structures also indicate the efficiency of the proposed definition, showing greater correlation coefficients between motion duration and the damages witnessed. To further clarify on this matter, the outcomes calculated for a 5-story structure (the model 1005), is presented in Table 4 as well.

Table 4 Correlation coefficients of response parameters with the record-based duration metrics for a 5-story structure.

| Response parameter | Correlation Coefficients | | | | | | |
|---|---|---|---|---|---|---|---|
| | Proposed definition | Bracketed 0.05g | Bracketed 0.1g | Uniform 0.05g | Uniform 0.1g | $D_{5-75}$ | $D_{5-95}$ |
| $DI_{total}$ | 0.916 | 0.742 | 0.766 | 0.817 | 0.775 | 0.761 | 0.737 |
| $DI_{s-1}$ (1st story) | 0.890 | 0.718 | 0.741 | 0.791 | 0.747 | 0.737 | 0.711 |
| $DI_m$ (beam) | 0.864 | 0.767 | 0.785 | 0.816 | 0.756 | 0.753 | 0.756 |
| $DI_m$ (column) | 0.892 | 0.704 | 0.727 | 0.784 | 0.744 | 0.732 | 0.698 |
| $EH_{total}$ | 0.943 | 0.760 | 0.788 | 0.854 | 0.827 | 0.804 | 0.754 |
| $EH_m$ (beam) | 0.933 | 0.778 | 0.784 | 0.827 | 0.768 | 0.766 | 0.774 |
| $EH_m$ (column) | 0.911 | 0.684 | 0.706 | 0.770 | 0.741 | 0.723 | 0.680 |

Two response-based definitions for motion duration, namely the ones recommended by Xie and Zhange (1988) as well as Zahrah and Hall (1984), are also considered in this study for the comparison purposes. Xie and Zhange (1988) put forward a duration definition—also called engineering duration (ENG-D)—in which the threshold of a uniform duration (e.g. 0.05g or 0.1g) is substituted by an engineered or structural-related formula, $F_y/(M \times BETA(T))$. With this formula, a new threshold can be found for a specific structure using its total mass ($M$), yield strength level ($F_y$) and $BETA(T)$. The term $BETA(T)$ is the ordinate of a selected earthquake response spectrum calculated for the desired damping ratio at the natural period (T) of the structure being studied. Besides, Zahrah and Hall (1984) used a function between cumulative hysteretic energy against the time (t) of a motion to recommend a response-based duration definition. This definition is also known as an effective duration (EFF-D) and is the length of time ($t_{e0.75} - t_{e0.05}$) within which a 5% up to a 75% of the earthquake energy in a structure is inelastically imparted.

The correlations of the structural response parameters, the damage metrics used in this paper, and two response-based definitions employed here have been investigated using the case study RC buildings modeled in this research. Results demonstrate that the proposed definition (the inelastic duration or IDU) gives improved correlation values compared to the ones obtained from the other response-based definitions, namely the effective and engineering duration of strong motions. In this case, the outcomes for models 1003 and 1008, respectively, are presented in Table 5 and Table 6, explaining more details pertaining to the efficiency of the proposed definition for earthquake duration.

Table 5 Correlation coefficients of response parameters with the response-based duration metrics for a 3-story structure.

| Response parameter | Correlation Coefficients | | |
| --- | --- | --- | --- |
| | Proposed definition (IDU) | Effective duration $(t_{e0.75} - t_{e0.05})$ | Engineering duration |
| DI$_{total}$ | 0.928 | 0.831 | 0.784 |
| DI$_{s-1}$ (1$^{st}$ story) | 0.914 | 0.828 | 0.781 |
| DI$_m$ (beam) | 0.925 | 0.826 | 0.873 |
| DI$_m$ (column) | 0.902 | 0.822 | 0.764 |
| EH$_{total}$ | 0.949 | 0.855 | 0.837 |
| EH$_m$ (beam) | 0.939 | 0.823 | 0.872 |
| EH$_m$ (column) | 0.917 | 0.821 | 0.759 |

Table 6 Correlation coefficients of response parameters with the response-based duration metrics for an 8-story structure.

| Response parameter | Correlation Coefficients | | |
| --- | --- | --- | --- |
| | Proposed definition (IDU) | Effective duration $(t_{e0.75} - t_{e0.05})$ | Engineering duration |
| DI$_{total}$ | 0.843 | 0.801 | 0.836 |
| DI$_{s-1}$ (1$^{st}$ story) | 0.812 | 0.750 | 0.750 |
| DI$_m$ (beam) | 0.742 | 0.676 | 0.678 |
| DI$_m$ (column) | 0.735 | 0.686 | 0.735 |
| EH$_{total}$ | 0.866 | 0.858 | 0.819 |
| EH$_m$ (beam) | 0.892 | 0.832 | 0.865 |
| EH$_m$ (column) | 0.886 | 0.831 | 0.870 |

In order to reach a conclusive result, average correlation coefficients for each response parameter are calculated using the outputs of all considered structures (models). In this case, all individual employed duration-related metrics, the proposed parameter (the IDU) plus other duration definitions, are incorporated to configure a ranking trend. Figure 10 shows the aforementioned average correlation coefficients of Park-Ang damage indices—for a beam, a story and the whole structural system—with different considered duration definitions. Compared with the other existing definitions, this figure shows that the proposed definition (inelastic duration or IDU) has an upper correlation coefficient (more than 80%) with all taken Park-Ang damage indices. The same procedure is performed for the hysteretic energy of two members (a beam and a column at the ground floor level) and the relevant energy term for the whole structural system. The results associated with the average correlation coefficients obtained for the aforementioned energy-based damage metrics are shown in Figure 11 (a) to (c), respectively. It is interesting to report that the IDU has a higher correlation with the employed energy-based damage indices, exceeding correlation coefficients greater than 0.9.

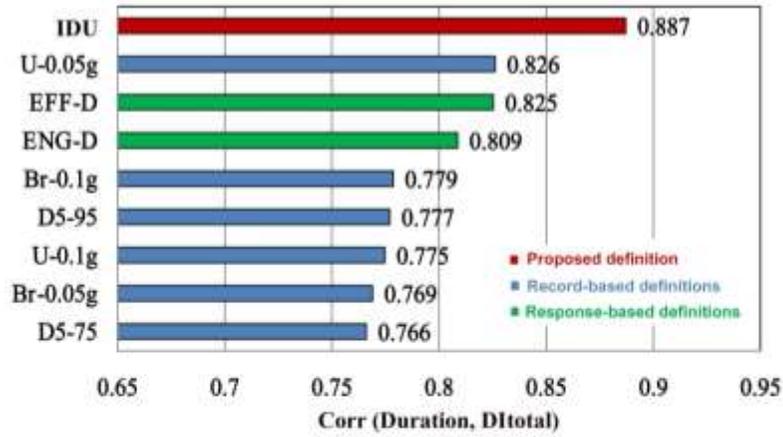
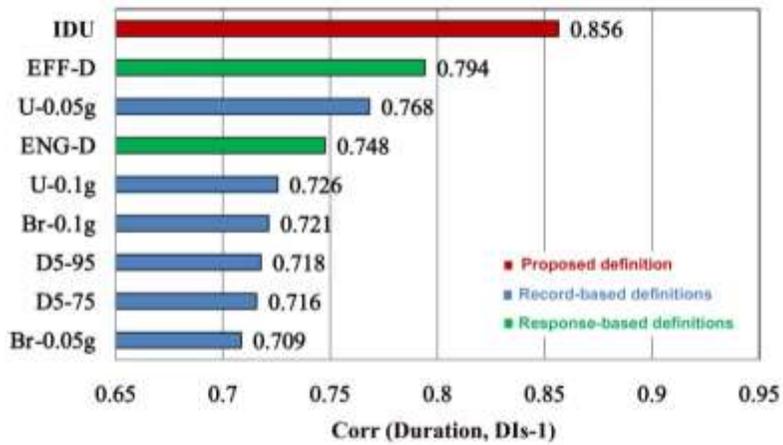
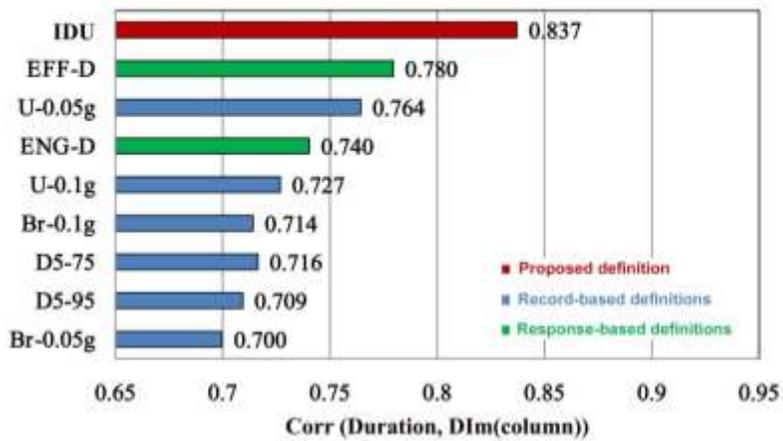

Figure 10. Average correlation coefficients of Park-Ang damage index and motion duration for the proposed method (IDU) and other duration definitions: a) total Park-Ang damage; b) Park-Ang damage for ground floor; c) Park-Ang damage for a column at the first story

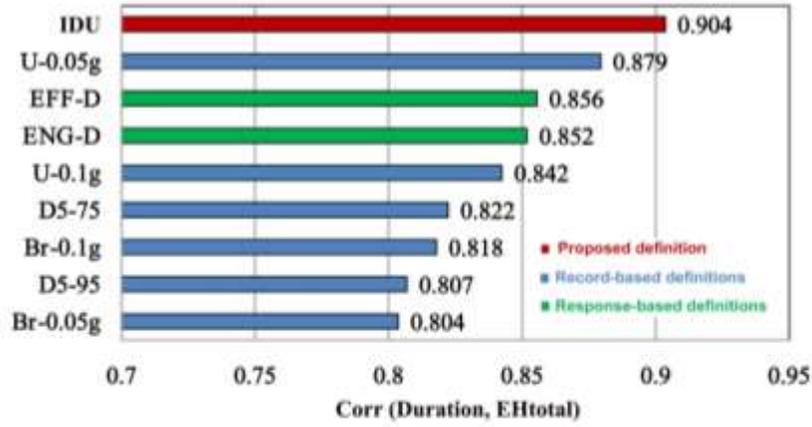

(a)

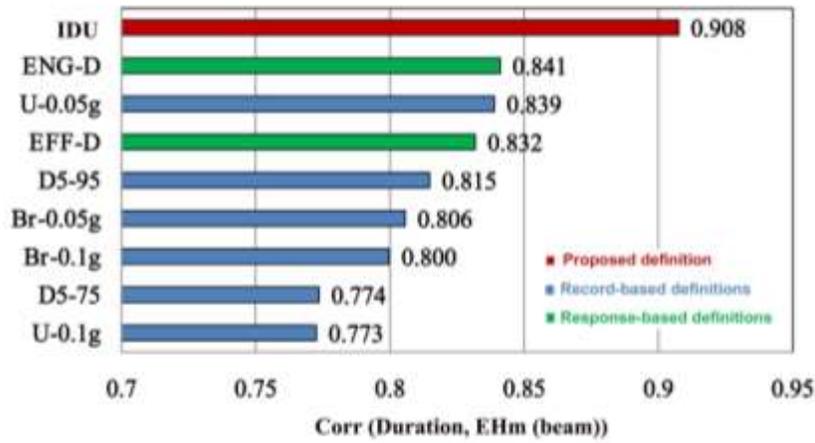

(b)

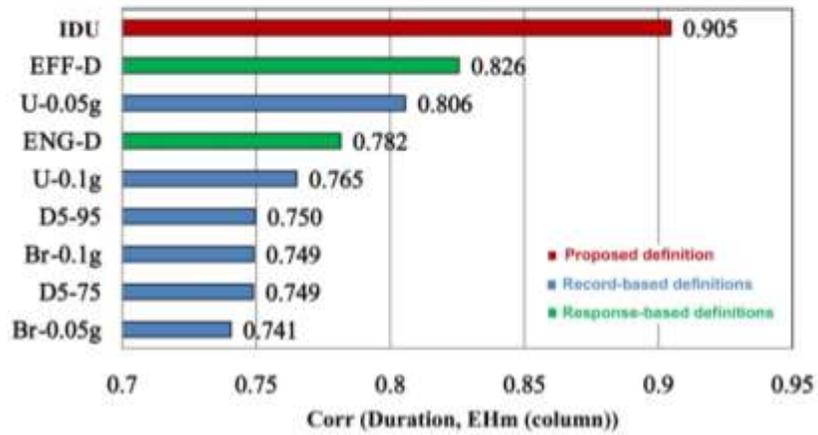

(c)

Figure 11. Average correlation coefficients of energy-based metrics and motion duration for the proposed method (IDU) and other duration definitions: a) total energy of the structures; and hysteretic energy of a beam (a) and a column (b) at the ground floor.

It can be seen that the response-based definitions including inelastic duration (or IDU) are located at the upper parts of the rankings displayed in figures 10 and 11. So, the hypothesis that they are more accurate than the record-based metrics to show the possible correlation of duration and structural damages seems to be correct. If we exclude the response-based definitions from this ranking, the uniform duration with a threshold of 0.05g demonstrates the best correlation with measured damages—the Park-Ang and hysteretic energy indices. In other words, the 0.05g uniform duration definition always ranks first in all cases investigated if record-based definitions are only considered for comparative evaluation. This finding is consistent with the results reported before (Guo et al. 2018; Hancock and Bommer 2007). However, Figures 10 and 11 illustrate that the second place in this ranking trend is interchangeably occupied by the other record-based definitions.

## 7   Discussion

In this section, various engineering applications of the proposed duration definition, inelastic duration, with a discussion of its pros and cons are presented. Whereas it has been already demonstrated that the inelastic duration reflects the influence of motion duration reliably, it should be clarified how this definition can be applied in engineering problems. This matter is investigated in three main problems in which motion duration may be properly incorporated. These three problems, including duration sensitivity analysis, structural seismic response assessment and structural reliability analysis, are respectively examined as follow:

- In the duration sensitivity analysis, the sensitivity extent of a structural type model to motion duration is investigated. The output of such an analysis is employed to determine whether or not motion duration must be incorporated in seismic analysis of a structural model type. Because this analysis is performed for a specific structure and its dynamic characteristics are achievable, the proposed definition can be readily applied. Using the inelastic duration in motion duration sensitivity analysis may be preferred because this duration metric, as revealed before, can better reflect the influence of earthquake duration if it is compared to the other studied definitions.

- In duration-consistent structural seismic response assessment, ground motions whose durations are compatible with a seismic hazard analysis of the region must be selected for time history dynamic analysis. In this case, appropriate ground motion prediction equation (GMPE) must be developed for seismic hazard analysis if the proposed duration metric is selected to be at work. In contrast to the present GMPEs—developed for record-based duration definitions—which are only a function of earthquake parameters such as magnitude, distance, shear wave velocity, GMPEs of the inelastic duration would be a function of both structural and earthquake parameters. It is worth mentioning that the functional forms of GMPEs on the inelastic duration can be somewhat similar to those derived for record-based durations with a difference that for the inelastic duration, constant parameters of the possible GMPE should be separately determined for each structure. Similar to those GMPEs developed for spectral acceleration in which constant parameters are evaluated for several SDOFs with a period ranging from 0-10sec, GMPEs that are derived to predict the inelastic duration can be developed for several SDOFs with periods ranging from 0-5sec and ductility ratios ranging from 2 to 12. Although the derivation of these GMPEs are very time-consuming and requires huge numerical analysis, engineers would only use the final developed models without any further consideration.

- The inelastic duration can be also used in structural reliability analysis (e.g. Mahsuli 2012) as an explanatory variable for predicting structural responses. If duration sensitivity analysis shows that motion duration should be incorporated in seismic response analysis of a structure, a modifier function can be developed to reflect the influence of motion duration on the predicted responses. This modifier function can be derived for a specific structure or for a structural model type. In the latter case, characteristics of structures such as heights and bays along with inelastic duration should be considered as variables in the function. In both cases, the inelastic duration can be used because structural-related parameters which are necessary for determining inelastic duration are available.

# 8 Conclusion

This study introduces a new definition for ground motion duration, which is based on the nonlinear response of a particular structure. The correlation coefficient of structural damage with motion duration is employed to explore the efficiency of the proposed definition. In this case, damages occurred in several reinforced concrete structures, subjected to spectrum matched ground motions, are considered for the correlation computation. Matching the spectrum of ground motions implies that the variability of structural responses is mainly attributed to the duration characteristics of ground motions. This investigation is conducted for four concrete frame structures, where cumulative damages are considered as response parameters. Results are as following:

1. Compared to the other existing definitions employed in this study, the proposed definition brings a 10 up to 15 percent improvement in the correlation between motion duration and the applied damage measures. It is worth to mention that highest correlation between motion duration and the structural damages is witnessed in all considered structures.

2. With the proposed duration definition, at least an 80% correlation coefficient between motion duration and the Park-Ang damage indices is observed—either for a local or a global scale. This emphasizes the high correlation between earthquake duration and damages imposed on the structures during strong shaking.

3. The proposed definition produces more than 90% correlation coefficients between motion duration and the hysteretic energy of the earthquakes. This illustrates that hysteretic cyclic characteristics of the structural members may get severely affected by long-duration earthquakes.

4. Among the record-based definitions of motion duration, the 0.05g uniform duration ranks first (if record-based metrics are only considered) almost for all damage measures applied in this study—Park-Ang and energy-related indices. This indicates that the record-based definitions that have a meaningful relationship with the structural damage are more successful in predicting the influence of duration length on the response of the built infrastructures.

## 9 Acknowledgment

The authors do thank Professor Julian J Bommer and another anonymous reviewer for meticulous review comments on an earlier version of this article. The authors would also like to thank all the efforts accomplished by the staffs working in the center of High-Performance Computing (HPC) at Sharif University of Technology for providing a reliable and fast platform for the required computational analyses of this project. It is worth mentioning that the first two authors of this paper have contributed equally to the work

Korkmaz A. Aktaş E. (2006). Probability based seismic analysis for r/c frame structures. *Journal of the Faculty of Engineering and Architecture of Gazi University*, *21*(1), 55–64.

Mahsuli, M. (2012). *Probabilistic models, methods, and software for evaluating risk to civil infrastructure*. *PhD THESIS*. THE UNIVERSITY OF BRITISH COLUMBIA.

Mashayekhi, M., & Estekanchi, H. . E. (2013). Investigation of strong-motion duration consistency in endurance time excitation functions. *Scientia Iranica*, *20*(4), 1085–1093.

Mashayekhi, M., & Estekanchi, H. E. (2012). Significance of effective number of cycles in Endurance Time analysis. *Asian Journal of Civil Engineering (Building and Housing)*, *13*(5), 647–657.

Mashayekhi, M., Harati, M., & Estekanchi, H. E. (2019). Estimating the duration effects in structural responses by a new energy-cycle based parameter. *Submitted for Publication*.

McKenna, F. (2014). Open System for Earthquake Engineering Simulation (OpenSees) version 2.4. 4 MP [Software].

Miranda, E. ., & Bertero, V. V. (1994). Evaluation of Strength Reduction Factors for Earthquake-Resistant Design. *Earthquake Spectra*, *10*(2), 357–379. https://doi.org/10.1193/1.1585778

Nassar, A.A and Krawinkler, H. (1991). *Seismic demands for SDOF and MDOF systems*. *Blume report*.

Nathan M. Newmark. (1959). A Method of Computation for Structural Dynamics. *Journal of the Engineering Mechanics Division*, *85*(3), 67–94.

Pan, Y., Ventura, C. E., & Liam Finn, W. D. (2018). Effects of Ground Motion Duration on the Seismic Performance and Collapse Rate of Light-Frame Wood Houses. *Journal of Structural Engineering*, *144*(8), 4018112. https://doi.org/10.1061/(ASCE)ST.1943-541X.0002104.

Park, Y., & Ang, A. H. -S. (1985). Mechanistic Seismic Damage Model for Reinforced Concrete. *Journal of Structural Engineering*, *111*(4), 722–739. https://doi.org/10.1061/(ASCE)0733-9445(1985)111:4(722)

Park, Y. J., Ang, A. H., & Wen, Y. K. . (1987). Damage-Limiting Aseismic Design of Buildings. *Earthquake Spectra*, *3*(1), 1–26.

Priestley, M. J. N., & Grant, D. N. (2005). Viscous Damping in Seismic Design and Analysis. *Journal of Earthquake Engineering*, *9*(sup2), 229–255. https://doi.org/10.1142/s1363246905002365

Raghunandan, M., & Liel, A. B. (2013). Effect of ground motion duration on earthquake-induced structural collapse. *Structural Safety*, *41*, 119–133. https://doi.org/10.1016/j.strusafe.2012.12.002

Reinhorn, A., Roh, H., Sivaselvan, M., Kunnath, S. K., Valles, R., Madan, A., … Park, Y. J. (2009). *IDARC2D Version 7.0: A Program for the Inelastic Damage Analysis of Structues*. *Technical Report MCEER-09-0006*. https://doi.org/10.13140/RG.2.1.2518.8724